\documentclass[pra,aps,showpacs,preprint,floatfix]{revtex4}

\usepackage{graphicx}
\usepackage{amsmath}
\usepackage{amsfonts}
\usepackage{amssymb}

\begin{document}


\title{Conserving Quasiparticle Calculations for Small Metal Clusters}

\author{G.~Pal}
\email{pal@physik.uni-kl.de} \affiliation{Physics Department and
  Research Center OPTIMAS, Kaiserslautern University, P.O.Box 3049, 67653
Kaiserslautern, Germany}
\author{Y. Pavlyukh}
\altaffiliation{Now at Institut f\"{u}r Physik, Martin-Luther-Universit\"{a}t
Halle-Wittenberg, Heinrich-Damerow-Strasse 4, 06120 Halle, Germany}
\author{H.~C.~Schneider}
\email{hcsch@physik.uni-kl.de} \affiliation{Physics Department and
  Research Center OPTIMAS, Kaiserslautern University, P.O.Box 3049, 67653
Kaiserslautern, Germany}
\author{W.~H\"{u}bner}
\affiliation{Physics Department and
  Research Center OPTIMAS, Kaiserslautern University, P.O.Box 3049, 67653
Kaiserslautern, Germany}

\date{\today}

\begin{abstract}
 A novel approach for GW-based calculations of quasiparticle
properties for finite systems is presented, in which the screened
interaction is obtained directly from a linear response calculation
of the density-density correlation function. The conserving nature
of our results is shown by explicit evaluation of the $f$-sum rule.
As  an application, energy renormalizations and level broadenings
are calculated for the closed-shell Na$_9^+$ and
Na$_{21}^+$ clusters, as well as for  Na$_4$. Pronounced improvements
of conserving approximations to RPA-level results are obtained.
\end{abstract}


\maketitle

\section{Introduction}

The GW approximation for the self-energy~\cite{Hedinfirst} is used
widely and with considerable
success~\cite{OnidaReiningRubio,Gunnarsson,EcheniqueAryasetiawan,%
Rohlfing2,HolmBarth,SchoeneEguiluz,Faleev,Kotliar} in Green's
function based methods to compute bulk and surface electronic band
structures and lifetimes for a variety of metals and semiconductors.
Within this approach, the self-energy $\Sigma=i\hbar GW$, computed
from the single-particle Green's function $G$ and the dynamically
screened interaction $W$, is the fundamental quantity that describes
the influence of electronic correlations on the single-particle band
structure. Despite the remarkable success of the standard GW theory,
which constructs $W$ from the dielectric function in the
random-phase approximation
(GW-RPA)~\cite{OnidaReiningRubio,Gunnarsson,EcheniqueAryasetiawan,Rohlfing2},
it is still debated  how self-consistency  should be implemented and
whether it is actually
needed~\cite{HolmBarth,SchoeneEguiluz,Faleev,Kotliar,Ku}.
Furthermore, there have been efforts to go beyond the GW
approximation by introducing vertex
corrections~\cite{MahanSernelius,Schindlmayr1998vertex,TammeSchepe,Takada}.

In this paper, we present a general approach for GW calculations of
quasiparticle (QP) properties, in which an accurate screened Coulomb
potential is calculated based on consistency requirements between
the single-particle Green's function (determined by the self-energy)
and the screened potential (determined by the density-density
correlation function). Such an approach is particularly well suited
for metal clusters, which combine features of finite systems with
those of extended
ones~\cite{BennemannPRL,EcheniqueNano,YaroslavHubner}. In clusters,
both correlations between electrons in localized states \emph{and}
collective excitations influence the band structure and therefore
must be taken into account. Usual ap\-prox\-i\-ma\-tions com\-mon to
GW calculations for the limiting cases of small atomic systems and
extended systems encounter potential problems for clusters. For
instance, it is difficult to describe screening effects in a quantum
chemistry approach based on perturbation theory starting from the
Hartree-Fock (HF) Hamiltonian. On the other hand, the influence of
electronic correlations cannot be captured by mean-field screening
models such as simple plasmon-pole or static approximations.
Therefore it is important to compute the dynamically screened
Coulomb potential, which contains two-particle correlations, as
accurately as possible in the course of the GW calculation.

The calculation of the screened Coulomb potential is the main
difference of our approach from standard GW-RPA and its selectively
improved versions. The motivation for this change in approach is
that $W$ as computed in the GW-RPA is an auxiliary quantity, which
lacks some of the properties of a physical screened
potential~\cite{HolmBarth}. To be more specific, GW-RPA does not
obey the charge/current conservation law as it applies to the
density-density correlation function, unless the screened potential
$W$ is calculated from single-particle Green's function in the
Hartree approximation~\cite{Baym-Kadanoff:PR}. This approximation,
however, is clear\-ly not good enough for clusters. Using Green's
functions determined self-consistently instead of the Hartree
Green's function seems to be a straightforward improvement for the
calculation of $W$, but it leads to a violation of the $f$-sum
rule~\cite{Szymansky}, and therefore is in conflict with our goal of
improving the quality of the screened Coulomb potential.

To obtain a screened Coulomb potential that does not violate sum
rules, we choose the self-energy first, and obtain the
\emph{consistent} screened potential (including vertex corrections) by
the physical constraint that the polarization function (or
density-density correlation function) fulfils charge/current
conservation~\cite{StrinatiMattauschHanke,Baym-Kadanoff:PR}. This
can be achieved by determining the density-density correlation
function as a functional derivative of the Green's function with
respect to an external
potential~\cite{Baym-Kadanoff:PR,Kwong-Bonitz}. Using the functional
derivative technique, the concept of $\Phi$-derivability is not
explicitly needed~\cite{BaymPR}, if one starts from conserving
Green's functions, as is the case in our calculations. Since we
cannot exactly compute the functional derivative of a Green's
function with respect to an external potential, we obtain an
approximate functional derivative by solving a linearized
quantum-kinetic equation for the one-particle Green's function in
the presence of a weak external potential with a generalized QP
ansatz. Particle-number conservation on the two-particle level is
explicitly checked by evaluating the $f$-sum rule.

Our GW approach for finite systems starts from the HF
single-particle energies and wavefunctions, because the correlations
described by the GW self-energy can be naturally divided into a HF
and a correlation part. The HF contribution is expected to be
important in finite systems, and our approach directly allows one to
introduce approximations in the correlation part, while keeping the
HF contribution unchanged.

At this point a few remarks are in order to relate our approach to
other strategies to perform GW calculations. Written in an abstract
form~\cite{Hedin-Lundqvist}, the interacting Green's function is
determined by the Dyson equation
\begin{equation}
G = G_0 + G_0 \Sigma G, \label{GDyson}
\end{equation}
where $G_0$ denotes the non-interacting Green's function. The
quality of a Green's function calculation is controlled by the self
energy
\begin{equation}
\Sigma =  i\hbar GW\Gamma \label{formal:sigma}
\end{equation}
which is given in terms of the screened potential
\begin{equation}
W =  v + v P W, \label{WDyson}
\end{equation}
with the bare Coulomb potential $v$ and the irreducible polarization
function
\begin{equation}
P = -i\hbar G G \Gamma. \label{formal:polarization}
\end{equation}
The irreducible polarization function can, in principle,  be
computed from the (two-particle) density-density correlation
function $\chi$, which is identical to the reducible polarization
function. The expressions~\eqref{formal:sigma}
and~\eqref{formal:polarization} contain the \emph{exact} vertex
function $\Gamma$
\begin{equation}
\Gamma= 1+\frac{\delta \Sigma}{\delta G} G G \Gamma,
\label{formal:vertex}
\end{equation}
which, in turn, depends on the \emph{exact} self energy $\Sigma$.
For approximations to the coupled set of equations
(\ref{GDyson})--(\ref{formal:vertex}), the vertex insertions for
self energy and polarization function need not
agree~\cite{Szymansky,StrinatiMattauschHanke}. An important question
for the design of a calculational procedure is therefore the
selection of approximations for the vertex corrections entering
Eqs.~\eqref{formal:sigma} and~\eqref{formal:polarization} together
with the choice of which of the quantities are to be updated in a
computational self-consistency cycle. For this choice, no criteria
exist \emph{in the framework of formal Green's functions
theory}~\cite{StrinatiMattauschHanke}, so that, for instance, the
calculational procedure can be determined by optimization for a
particular system and/or the physical quantities of
interest~\cite{Almbladh}. Or the vertex corrections and the
self-consistency procedure are chosen using a ``best $G$, best $W$''
philosophy, which aims at an optimization of the two quantities
separately, although it has been shown that this approach should be
avoided~\cite{DelSolePRB}. A different approach to choosing the
vertex corrections is based on diagrammatic arguments and has been
applied to extended
systems~\cite{MahanSernelius,Schindlmayr1998vertex} and finite
systems~\cite{ShirleyMartin}. The latter reference uses, for
instance, the same exchange vertex both in the self energy and the
polarization function. Using diagrammatic arguments, extra care
needs to be taken in order to avoid double counting of diagrams in
the expressions for the self energy and the polarization.

Instead of using any of the above mentioned techniques to design
suitable approximations, one can also use \emph{general} criteria
derived from physical conservation laws or invariance principles.
Such an approach has been used for transport
calculations~\cite{Baym-Kadanoff:PR} and electronic structure
calculations~\cite{StrinatiMattauschHanke,Szymansky,Takada}. Since
the conservation laws imply important consistency conditions for the
one and two-particle correlation
functions~\cite{Baym-Kadanoff:PR,StrinatiMattauschHanke}, they
prevent a choice of the vertex contributions that optimizes, say,
the one-particle properties at the cost of the two-particle
correlation functions. The choice of vertex corrections, which are
inconsistent in the sense that they do not obey the charge-current
conservation law, may be suitable for a particular calculation, but
such a choice is more likely to encounter problems with calculations
where both the single-particle quantities (i.e., $G$ and $\Sigma$)
and the two-particle quantities (i.e., $W$ and $P$) are intimately
connected. The question of consistency between the different
``ingredients'' for the calculation also arises if one uses, for
instance, density-functional based single-particle states as input
in GW calculations, which are then used as input to Bethe-Salpeter
equation calculations~\cite{OnidaReiningRubio}.

The manuscript is organized as follows. The theoretical approach is
described in Section~\ref{Section2}. Section~\ref{Section3}
presents, as an application of the general method, results on the
renormalized single-particle energies and level broadenings in Na
clusters of different sizes. The $f$-sum rule for finite systems is
derived in Appendix~\ref{AppendixA}. To facilitate the comparison of
our approach to existing calculations based on the Bethe-Salpeter
equation, a comparison between the two is presented in
Appendix~\ref{AppendixB}. Appendix~\ref{AppendixB} also demonstrates
the relation of our quantum-kinetic calculation in
Sec.~\ref{Section2} with the work of Baym and
Kadanoff~\cite{Baym-Kadanoff:PR}.

\section{Theory}\label{Section2}

\subsection{Equilibrium relations}

We start by introducing our notations and by presenting the
necessary equations for the equilibrium Green's functions formalism
for finite systems that is used to formulate the GW theory. For the
calculation of the one-particle Green's function from the
dynamically screened potential we use the GW self-energy
\begin{equation}
\Sigma(1,2) = i\hbar G(1,2)W(2,1),
\end{equation}
where $1=(\mathbf{r}_1,\sigma_1,t_1)$ denotes the space, spin and
time variable. The spatial dependence of the functions is expanded
in a basis of HF eigenfunctions $\{\varphi_{n}({\mathbf r})\}$,
where $n$ labels the HF spin orbital. For single-particle quantities
such as $G$ and $\Sigma$, we employ the matrix notation
\begin{equation}\label{Gnn}
G_{n_1n_2}(t_1,t_2)=\sum_{\sigma_1 \sigma_2}\int
\varphi^*_{n_1}({\mathbf r}_1) G(1,2) \varphi_{n_2}({\mathbf
r}_2){\mathrm d}^3r_1{\mathrm d}^3r_2,
\end{equation}
while for two-particle quantities, such as the bare ($v$) or the
screened ($W$) Coulomb potentials, we use
\begin{equation}\label{indexCoulomb}
\langle n_1n_2 | W(t_1,t_2) |n_3n_4 \rangle  =\sum_{\sigma_1 \dots
\sigma_4}
\int\varphi^*_{n_1}({\mathbf r}_1) \varphi^*_{n_2}({\mathbf r}_2)
  W(1,2) \varphi_{n_3}({\mathbf r}_1) \varphi_{n_4}({\mathbf
r}_2){\mathrm d}^3r_1{\mathrm d}^3r_2 .
\end{equation}

The Dyson equation for the equilibrium ($T=0\,\mathrm{K}$) retarded
Green's function~\cite{Hedin-Lundqvist}
\begin{equation}\label{Dyson}
\sum_{n_2}  \left[\hbar(\omega + i\gamma)\delta_{n_1n_2}
-\Sigma^{\mathrm{HF}}_{n_1n_2}-\Sigma^{\mathrm{corr}}_{n_1n_2}(\omega)\right]
G^{\mathrm{r}}_{n_2n_3}(\omega)= \delta_{n_1n_3}
\end{equation}
can be written in a form which explicitly displays the static HF and
dynamic correlation contributions to the retarded self-energy. The
HF self energy
\begin{equation}
\label{sigma-HF} \Sigma^{\mathrm{HF}}_{n_1n_2}= \sum_{n_3n_4} \bigl[
\langle n_1n_3 | v|n_2n_4 \rangle -\langle n_1n_2 | v|n_3n_4
\rangle\bigr]f_{n_3n_4}
\end{equation}
is determined by the direct and exchange Coulomb matrix elements as
well as the one-particle distribution functions
\begin{equation}
\label{one-particle-rho} f_{n_1n_2}=-i\hbar \int\frac{d\omega}{2\pi}
G^<_{n_1n_2}(\omega).
\end{equation}
The correlation contribution to the retarded self-energy
$\Sigma^{\mathrm{corr}}(\omega) = \Sigma^{\mathrm{r}}(\omega) -
\Sigma^{\mathrm{HF}}$ is connected to the lesser and the greater
components by the identity
\begin{equation}\label{sigmacorr}
\mathrm{Im}\Sigma^{\mathrm{corr}}_{n_1n_1}(\omega) =
\frac{1}{2i}\left( \Sigma^{>}_{n_1n_1}(\omega)-
\Sigma^{<}_{n_1n_1}(\omega) \right).
\end{equation}
In the GW approximation and in a discrete basis function
representation, they read
\begin{equation}\label{GWapprox}
\Sigma^{\gtrless}_{n_1n_1}(\omega)= i\hbar \sum_{n_2} \int
\frac{{\mathrm d}\omega'}{2\pi} 
G^{\gtrless}_{n_2n_2}(\omega+\omega') \langle n_1n_1 |
W^{\lessgtr}(\omega') |n_2n_2 \rangle.
\end{equation}
We now need to relate the $>$ and $<$ components of $G$ and $W$ to
the retarded functions. For this, we use the fermionic
Kubo-Martin-Schwinger conditions~\cite{Kadanoff-Baym:Book}
\begin{align}\label{KMS_G}
G^<(\omega)=& -2\,i n_{\mathrm{F}}(\omega)\, \mathrm{Im}G^{\mathrm r}(\omega) \\
G^>(\omega)=& -2\,i \big[n_{\mathrm{F}}(\omega)-1\big] \,
\mathrm{Im}G^{\mathrm r}(\omega) ,
\end{align}
with the Fermi function
$n_{\mathrm{F}}(\omega)=[\exp(\beta\hbar\omega)+1]^{-1}$, where
$\beta=1/k_BT$. In the $T=0$\,K limit,
$n_{\mathrm{F}}(\omega)=\Theta(\epsilon_{\mathrm F}-\hbar\omega)$,
with the step function $\Theta$ and $\epsilon_{\mathrm F}$ the Fermi
energy. Then one obtains for Eq.~\eqref{one-particle-rho}, which
enters the HF self energy,
\begin{equation}
\label{def-fnm}
f_{n_1n_2}=\frac{\hbar}{\pi}\int_{-\infty}^{\epsilon_{\mathrm F}}
d\omega \mathrm{Im}G^{\mathrm{r}}_{n_1n_2}(\omega).
\end{equation}

For the screened Coulomb potential, which is related to a
two-particle correlation function, the bosonic Kubo-Martin-Schwinger
conditions read~\cite{Kuznetsov}
\begin{align}\label{KMS_W}
W^<(\omega)=& \, 2\,i n_{\mathrm{B}}(\omega)\, \mathrm{Im}W^{\mathrm r}(\omega) \\
W^>(\omega)=& \, 2 \, i \big(n_{\mathrm{B}}(\omega)+1\big) \,
\mathrm{Im}W^{\mathrm r}(\omega).
\end{align}
In Eq.~\eqref{KMS_W}, $n_{\mathrm{B}}(\omega)=
[\exp(\beta\hbar\omega)-1]^{-1}$ is the Bose function. In the
$T=0$\,K limit, $n_{\mathrm{B}}(\omega)=-\Theta(-\hbar \omega)$.
Inserting~\eqref{KMS_G} and~\eqref{KMS_W} into~\eqref{GWapprox}, we
obtain for the self-energies
\begin{align}\label{Sigma_lessgtr}
\Sigma^{>}_{n_1n_1}(\omega)& =  4i\hbar  \sum_{n_2} \int
\frac{{\mathrm d}\omega'}{2\pi}  
n_{\mathrm{B}}(\omega')\big[ n_{\mathrm{F}}(\omega+\omega')-1\big] 
 \mathrm{Im} G^{\mathrm r}_{n_2n_2}(\omega+\omega') \,
\mathrm{Im}\langle n_1n_1 | W^{\mathrm r}(\omega') |n_2n_2 \rangle,\\
\Sigma^{<}_{n_1n_1}(\omega)&=  4i\hbar  \sum_{n_2} \int
\frac{{\mathrm d}\omega'}{2\pi} \big[ n_{\mathrm{B}}(\omega')+1 \big]
n_{\mathrm{F}}(\omega+\omega')  \mathrm{Im} G^{\mathrm r}_{n_2n_2}(\omega+\omega') \,
\mathrm{Im}\langle n_1n_1 | W^{\mathrm r}(\omega') |n_2n_2 \rangle .
\end{align}
From Eqs.~\eqref{sigmacorr} and~\eqref{Sigma_lessgtr},  a
Montroll-Ward expression~\cite{KraeftPREetal} for the imaginary part
of the correlated (retarded) self-energy
\begin{equation}\label{ImSigma}
\mathrm{Im}\Sigma^{\mathrm{corr}}_{n_1n_1} (\omega)= \hbar
\sum_{n_2} \int \frac{{\mathrm d}\omega'}{\pi} \left[
n_{\mathrm{B}}(\omega')+n_{\mathrm{F}}(\omega+\omega') \right]
 \mathrm{Im} G^{\mathrm{r}}_{n_2n_2}(\omega+\omega') \,
\mathrm{Im} \langle n_1n_1 | W^{\mathrm{r}}(\omega') |n_2n_2 \rangle
\end{equation}
can be derived. The real part of $\Sigma^{\mathrm{corr}}(\omega)$ is
calculated from a Kramers-Kronig transformation of
Eq.~\eqref{ImSigma}.

The computation of $\Sigma^{\mathrm{corr}}$ is closely linked with
the screened Coulomb potential, $W^{\mathrm{r}}$ or, equivalently,
the inverse dielectric function. They can be con\-structed from the
retarded density-density correlation function $\chi^{\mathrm{r}}$
using
\begin{equation}\label{Weps}
W^{ \mathrm{r}}(1,2)  = \int {\mathrm d} 3 \varepsilon^{-1}(1,3)\, v(3,2)
 =v(1,2)+\int {\mathrm d} 3 {\mathrm d} 4 \, v(1,3)
\chi^{\mathrm{r}}(3,4) \, v(4,2), 
\end{equation}
where $\int d3= \int d^3r_3 dt_3$. The correlation function
$\chi^{\mathrm r}$ is also the retarded density-response
function~\cite{Baym-Kadanoff:PR,Hedin-Lundqvist,Kwong-Bonitz} with
respect to a weak external perturbation $U$, i.e.,
\begin{equation}\label{definition:chi}
\chi^{ \mathrm{r}}(1,2)= \left.\frac{\delta \langle \rho (1)\rangle
}{\delta U(2 )}\right|_{U=0},
\end{equation}
where $\rho(1)=\psi^{\dag}(1)\psi(1)$ is the particle density
operator, expressed through the creation and destruction field
operators $\psi^{\dag}$ and $\psi$, respectively. The response
function in Eq.~\eqref{definition:chi} can therefore be calculated
in the framework of nonequilibrium Green's
functions~\cite{Binder-Koch,Kremp:Book,Kwong-Bonitz}, using exactly
the same GW approximation as employed for the determination of the
single-quasiparticle properties. The determination of a dielectric
function that is consistent with the GW self-energy is carried out
in the next section.

\subsection{Quantum kinetics}

In our basis function representation, the single-particle density
matrix is given in terms of a single-particle non\-equilibrium
Green's function
\begin{equation}
\label{GFdefinition} i\hbar G_{n_2n_1}^{<}(t,t)= -\langle
\rho_{n_1n_2}(t)\rangle = - \langle c^{\dag}_{n_1}(t) c_{n_2}(t)
\rangle,
\end{equation}
where $c^{\dag}_n$ and $c_n$ are the creation and the annihilation
operators of particles in the molecular orbital $n$, respectively.

In the following, we relate the determination of  $\chi^{\mathrm r}$
to a quantum-kinetic calculation of a nonequilibrium Green's
function under the action of a weak external field $U$.  For this,
we start from the Hamiltonian
\begin{align}
H= & \sum_{n_1}T_{n_1n_1}c^{\dag}_{n_1}(t)c_{n_1}(t) +
\sum_{n_1 n_2}U_{n_1n_2}(t) c^{\dag}_{n_1}(t)c_{n_2}(t) \nonumber \\
& +  \frac{1}{2}\sum_{n_1\dots n_4} \langle n_1n_2 | v |n_4n_3
\rangle c^{\dag}_{n_1}(t) c^{\dag}_{n_2}(t)c_{n_3}(t) c_{n_4}(t),
\end{align}
where $T$ is the kinetic part, which in our case includes the core
potential, and $v$ is the bare Coulomb matrix element, with the
index structure defined in Eq.~\eqref{indexCoulomb}. With this
Hamiltonian, the Green's function~\eqref{GFdefinition} evolves in
time according to
\begin{align}\label{eqmotGone}
i\hbar  \frac{\partial}{\partial t_1} G^<_{n_1n_2}(t_1,t_2)=\mbox{} &
\delta(t_1-t_2)\delta_{n_1n_2} \\ & + T_{n_1n_1}(t_1)
G^<_{n_1n_2}(t_1,t_2) + \sum_{n_3}U_{n_1n_3}(t_1)
G^<_{n_3n_2}(t_1,t_2) \nonumber \\
& + \sum_{n_3} \int \mathrm{d}t_3 \big( \Sigma^{\mathrm
r}_{n_1n_3}(t_1,t_3) G^<_{n_3n_2}(t_3,t_2)+ \Sigma^<_{n_1n_3}(t_1,t_3) G^{\mathrm
a}_{n_3n_2}(t_3,t_2)\big) . \nonumber
\end{align}
There is also the adjoint equation, corresponding to the derivative
with respect to $t_2$.

We next compile the relations between $G$, $W$ and $\Sigma$ that
generalize the equilibrium
functions~\eqref{sigmacorr}--\eqref{Sigma_lessgtr} to dynamical
quantities, and which are needed for the evaluation of
Eq.~\eqref{eqmotGone}
\begin{align}
G^{\mathrm r}(t_1,t_2)=\mbox{}& \Theta(t_1-t_2)\big( G^>(t_1, t_2)- G^<(t_1, t_2) \big) \\
W^{\mathrm r}(t_1,t_2)= \mbox{}& v\,\delta(t_1-t_2) +\Theta(t_1-t_2)
\big( W^>(t_1, t_2)- W^<(t_1, t_2) \big) \\
\Sigma^{\mathrm r}(t_1,t_2) = \mbox{} & \Sigma^{\mathrm{HF}}(t_1) \delta(t_1-t_2) 
+ \Theta(t_1-t_2)\big( \Sigma^>(t_1, t_2)-
\Sigma^<(t_1, t_2) \big). 
\end{align}
The instantaneous HF contribution is given by:
\begin{equation}
\Sigma^{\mathrm{HF}}_{n_1n_2}(t_1)=   -i\hbar  \sum_{n_3n_4}
\langle
n_1n_3 | v|n_2n_4 \rangle G^<_{n_3n_4}(t_1,t_1)
+ i\hbar  \sum_{n_3n_4} \langle n_1n_2 |v|n_3n_4 \rangle
G^<_{n_3n_4}(t_1,t_1).
\end{equation}
Finally, for $\Sigma^{\gtrless}(t_1,t_2)$ we use the GW form
\begin{equation}\label{GW_time}
\Sigma_{n_1n_2}^{\gtrless}(t_1, t_2)=  i\hbar \sum_{n_3n_4} 
G^{\gtrless}_{n_3n_4}(t_1, t_2) 
\langle n_1n_2 | W^{\lessgtr}(t_2, t_1)| n_3 n_4 \rangle.
\end{equation}

The above equations completely determine the system's re\-sponse to
the external potential in the GW approximation \emph{if} they are
supplemented by the dynamical equations for another Green's
function, say, for the retarded Green's function
$G^{\mathrm{r}}(t_1,t_2)$. In this scheme, both $G^{<}$ and
$G^{\mathrm{r}}$ depend on two time arguments, and the retarded
function describes the changes of the spectral properties of the
systems during the time evolution. An implementation of these
two-time equations for the electron gas was carried out in
Ref.~\cite{Kwong-Bonitz}. In this case, the numerical calculation is
run starting from non-interacting Green's functions without external
field for some time to obtain the interacting Green's functions,
which are then disturbed by the field. The dynamics under the
influence of the weak driving field then allows one to numerically
calculate the functional derivative, which determines the dielectric
function via Eq.~\eqref{definition:chi}. Since one can use the same
self-energy, say, in the GW approximation for both the kinetic and
the spectral Green's function, the dielectric function determined in
this way is consistent by construction with the single-particle
Green's function determined from the same self-energy.

Equation~\eqref{eqmotGone} for the dynamical Green's functions
depending on two real time arguments is an extremely complex
integro-differential equations, whose solution is possible only for
small or homogeneous systems~\cite{theDuchGuy}. For systems of
intermediate size, our aim is to develop a flexible approximate
numerical scheme which works only with Green's functions depending
on a single time argument. To this end, one can introduce
approximations, so that the resulting equations depend only on
$G^{\mathrm{r}}(t_1 - t_2)$, i.e., the equilibrium retarded Green's
function, whose Fourier transformation $G^{\mathrm{r}}(\omega)$ has
a simple physical interpretation, instead of
$G^{\mathrm{r}}(t_1,t_2)$. An important consequence of this
approximation is that the equilibrium $G^{\mathrm{r}}(t_1 - t_2)$
does not need to be calculated \emph{together} with the dynamical
Eq.~\eqref{eqmotGone}. Rather, the response of the system described
by Eq.~\eqref{eqmotGone} now becomes implicitly dependent on
$G^{\mathrm{r}}(\omega)$. Of course, the density-response function
determines the screening properties, so that
$G^{\mathrm{r}}(\omega)$ depends on the response calculation. This
interdependence introduces the possibility of a self-consistent
numerical procedure, in which one or both of these quantities are
updated and recalculated during the self-consistency cycle.

In this paper we concentrate on setting up the numerical procedure for
a consistent GW calculation in the spirit of Hedin's original GW
treatment as a ``one-shot'' self-energy correction to the HF ground
state. We do not intend to study the issue of self-consistency
here. We begin by writing down the Green's functions of the finite
system in the HF approximation
\begin{equation}\label{Dyson-HF}
\sum_{n_2}\left[\hbar(\omega + i\gamma)\delta_{n_1n_2}
-\Sigma^{\mathrm{HF}}_{n_1n_2}\right] G^{\mathrm{r\
(HF)}}_{n_2n_3}(\omega)= \delta_{n_1n_3}.
\end{equation}
Using HF spin orbitals as single-particle quantum numbers, this GF
becomes diagonal
\begin{equation}\label{Gr-HF}
G^{\mathrm{r\ (HF)}}_{n_1n_2}(\omega)=\frac{1}{\hbar(\omega +
i\gamma) -\epsilon^{\mathrm{HF}}_{n_1}} \ \delta_{n_1n_2} .
\end{equation}
In equilibrium, this corresponds also to a diagonal kinetic Green's
function
\begin{equation}
\label{G<-HF} G^{<\ \mathrm{(HF)}}_{n_1n_2}(\omega) = -2\,i
n_{\mathrm{F}}(\omega)\,\mathrm{Im}G^{\mathrm{r\
(HF)}}_{n_1n_1}(\omega)\  \delta_{n_1n_2} .
\end{equation}
In the limit $\gamma \to 0$, the equilibrium single-particle
correlation functions are simply given by
\begin{align}
\label{G<-HF2} G^{<\ \mathrm{(HF)}}_{n_1n_2} (\omega) & = 2\pi i
f_{n_1n_2}^{\mathrm{(HF)}}
\delta(\hbar\omega-\epsilon^{\mathrm{HF}}_{n_1})\\
f_{n_1n_2}^{\mathrm{(HF)}} &\mbox{} =
n_{\mathrm{F}}(\epsilon^{\mathrm{HF}}_{n_1})\,\delta_{n_1n_2},
\label{rho-HF}
\end{align}
Employing Eq.~\eqref{G<-HF2} as a description of the HF ground
state, we determine the dielectric function via
Eq.~\eqref{definition:chi}. To do this, we need several additional
steps and an approximation for the two-time kinetic Green's
function. We first note that, by definition of the functional
derivative, we only need to be interested in the \emph{linear
response} to a weak time-dependent perturbing potential $U$. To
first order in the weak perturbation, only density
\emph{fluctuations}, i.e., averages of the form $\langle
\rho_{n_1n_2}(t)\rangle$ with $n_1\neq n_2$, are driven away from
their equilibrium value~\eqref{rho-HF}, while the level occupations
$\langle \rho_{n_1n_1} \rangle $ remain equal to
$f_{n_1n_1}^{\mathrm{(HF)}}=
n_{\mathrm{F}}(\epsilon^{\mathrm{HF}}_{n_1})$ for all times. In the
numerical calculations, we take $\epsilon_{\mathrm F}$ to be in the
middle of the gap between the highest occupied and lowest unoccupied
molecular orbital.


Transcribing this back into the language of Green's functions using
Eq.~\eqref{GFdefinition}, we need to calculate the Green's functions
off-diagonal in the level indices $i\hbar G^{<}_{n_1n_2}(t,t)$, with
$n_1\neq n_2$. The quantum kinetic equation for these quantities is
derived by subtracting Eq.~\eqref{eqmotGone} and the adjoint
equation, making the substitution $t=(t_1+t_2)/2$ and $\tau=t_1-t_2$
and finally considering the equal-time limit $t_1=t_2=t$, $\tau=0$:
\begin{equation}\label{eqnmotG}
(i\hbar \frac{\partial}{\partial t}-\epsilon_{\alpha}^{\mathrm{HF}})
G^<_{\alpha}(t) + n_{\alpha}\Omega^{\mathrm{eff}}_{\alpha}(t)=
S_{\alpha}(t) .
\end{equation}
Here, $\alpha=(n_1,n_2)$ is a pair-state index for the off-diagonal
Green's function, $\epsilon_{\alpha}^{\mathrm{HF}} =
\epsilon_{n_1}^{\mathrm{HF}}-\epsilon_{n_2}^{\mathrm{HF}}$ the
energy difference between two levels, and
$n_{\alpha}=n_{\mathrm{F}}(\epsilon_{n_1})-n_{\mathrm{F}}(\epsilon_{n_2})$
is the difference in level distribution between the two
spin-orbitals of the pair state. The latter quantity is sometimes
referred to as the Pauli-blocking factor. The generalized driving
term
\begin{equation}\label{Rabi}
\hbar \Omega^{\mathrm{eff}}_{\alpha}(t)= i U_{\alpha}(t)+ \hbar
\sum_{\beta}(v^{\mathrm{ dir}}_{\alpha\beta} - v^{
\mathrm{exc}}_{\alpha\beta}) G^<_{\beta}(t)
\end{equation}
contains Coulomb enhancement contributions involving direct and
exchange matrix elements $v^{\mathrm{ dir}}$ and $v^{\mathrm{ exc}}$
\begin{align}
v_{\alpha\beta}^{\mathrm{ dir}}= v_{(n_1n_2)(n_3n_4)}^{\mathrm{
dir}}= \langle
n_1n_3 | v|n_2n_4 \rangle, \\
v_{\alpha\beta}^{\mathrm{ exc}}= v_{(n_1n_2)(n_3n_4)}^{\mathrm{
exc}}= \langle n_1n_2 | v|n_3n_4 \rangle.
\end{align}
The right hand side of Eq.~\eqref{eqnmotG} is the correlation term
\begin{equation}\label{Scat}
S_{n_1n_2}(t)= \sum_{n_3}\int_{-\infty}^{t}  {\mathrm{ d}}\bar t
\big[ \Sigma^>_{n_1n_3}(t, \bar t) G^<_{n_3n_2}(\bar t, t) +
G^<_{n_1n_3}(t, \bar t) \Sigma^>_{n_3n_2}(\bar t, t)  -(\lessgtr
\leftrightarrow \gtrless)\big]
\end{equation}
that accounts for interaction effects beyond HF. For the
self-energies, we use Eq.~\eqref{GW_time}.

The aim of our approach is to reduce the computational complexity of
the Green's functions depending on two time arguments, where kinetic
and spectral properties are tied closely together, by splitting the
problem into the the determination of the equilibrium $G^{\mathrm
  r}(\omega)$ from the calculation of the density-response function,
i.e., $G^{<}(t,t)$ in the presence of an external perturbation.
The main approximation involved in this split is that the two-time
Green's functions $G^{<}(t,\bar{t})$, which are contained in the
correlation contribution~\eqref{Scat} need to be related to the
dynamics of the density response, i.e., the time-diagonal Green's
function by virtue of Eq.~\eqref{GFdefinition}. To this end, we
employ a generalized Kadanoff-Baym ansatz in the
form~\cite{JahnkeKiraKoch}
\begin{equation}
G^{\gtrless}_{n_1n_2}(t , \bar t)= i\hbar
G_{n_1n_1}^{\mathrm{r}}(t- \bar t) G^{\gtrless}_{n_1n_2}(\bar t)
- i\hbar G^{\gtrless}_{n_1n_2}(t)
G_{n_2n_2}^{\mathrm{a}}(t-\bar t) \ .  
\end{equation}
For notational simplicity, we write here and in the following
$G^{\gtrless}_{n_1n_2}(t)$ for $G^{\gtrless}_{n_1n_2}(t,t)$.  To
evaluate the correlation contribution in the generalized
Kadanoff-Baym ansatz, the retarded $G^{\mathrm{r}}$ and advanced
$G^{\mathrm{a}}$ Green's functions, we employ the Hartree-Fock
Green's functions in the time domain:
\begin{align}
i\hbar G_{n_1n_1}^{\mathrm{r\ (HF)}}(t-\bar t)&= \Theta(t-\bar t)
\exp \{-\frac{i}{\hbar}
\tilde \epsilon_{n_1}(t-\bar t)\} \\
i\hbar G_{n_2n_2}^{\mathrm{a\ (HF)}}(\bar t-t)&\mbox{}= - \Theta(t-
\bar t) \exp \{ -\frac{i}{ \hbar}\tilde \epsilon^*_{n_2} (\bar
t-t)\}.
\end{align}
A non-zero ``background'' broadening $\gamma$ ensures the proper
behavior of the HF retarded and advanced Green's functions, and we
use the notation $\tilde\epsilon_n=
\epsilon^{\mathrm{HF}}_n-i\gamma$ . Then the correlation
contribution becomes
\begin{align}\label{S_interm1}
S_{n_1n_2}(t)=i\hbar \sum_{n_3n_4n_5} \int_{-\infty}^{t}d\bar t
\big[ &
e^{-\frac{i}{\hbar}(\tilde\epsilon_{n_4}-\tilde\epsilon_{n_2}^*)(t-\bar
t)} \langle n_1n_3 | W^{<}(\bar t,t) |n_4n_5 \rangle
G^<_{n_3n_2}(\bar t) G^>_{n_4n_5}(\bar t)           \\
+&e^{-\frac{i}{\hbar}(\tilde\epsilon_{n_1}-\tilde\epsilon_{n_5}^*)(t-\bar
t)} \langle n_3n_2 | W^{<}(t,\bar t) |n_4n_5 \rangle
G^<_{n_1n_3}(\bar t) G^>_{n_4n_5}(\bar t) \nonumber \\
& -  (\lessgtr \leftrightarrow \gtrless)\big]. \nonumber
\end{align}
Because we wish to determine the \emph{linear} density response to
the weak external perturbation, $S(t)$ is linearized with respect to
the off-diagonal $G^{\gtrless}$s that are driven by $U$. In the
spirit of linear response, the Green's functions appearing in one
term together with one off-diagonal Green's function are replaced by
the equilibrium relations $G^<_{n_1n_1}=\frac{1}{i\hbar}(1-f_{n_1})$
and $G^>_{n_1n_1}=-\frac{1}{i\hbar}f_{n_1}$, where we have defined
$f_{n_1}\equiv f^{\mathrm{(HF)}}_{n_1n_1}$. To further simplify the
equations, note also that $G^>_{n_1n_2}(\bar t)=G^<_{n_1n_2}(\bar
t)$, with $n_1 \neq n_2$. Equation~\eqref{S_interm1} then becomes
\begin{align}\label{S_interm2p}
S_{n_1n_2}(t)= \sum_{n_3n_4n_5} \int_{-\infty}^{t} {\mathrm{
d}}\bar t &
e^{-\frac{i}{\hbar}(\tilde\epsilon_{n_4}-\tilde\epsilon_{n_2}^*)(t-\bar
t)}  \\
\Big[ \delta_{n_4n_5} & G^<_{n_3n_2}(\bar t)\Big( \langle n_1n_3 |
W^{<}(\bar t, t) |n_4n_5 \rangle (1-f_{n_4}) + \langle n_1n_3 |
W^{>}(\bar t, t) |n_4n_5 \rangle f_{n_4} \Big) \nonumber \\
-\delta_{n_3n_2} & G^<_{n_4n_5}(\bar t)  \Big( \langle n_1n_3 |
W^{<}(\bar t, t) |n_4n_5 \rangle f_{n_2} + \langle n_1n_3 |
W^{>}(\bar t, t) |n_4n_5 \rangle (1-f_{n_2}) \Big) \Big] \nonumber \\
+ \sum_{n_3n_4n_5} \int_{-\infty}^{t}  {\mathrm{ d}}\bar t &
e^{-\frac{i}{\hbar}(\tilde\epsilon_{n_1}-\tilde\epsilon_{n_5}^*)(t-\bar
t)} \nonumber \\
\Big[ \delta_{n_4n_5} & G^<_{n_1n_3}(\bar t) \Big( \langle n_3n_2 |
W^{<}(t, \bar t) |n_4n_5 \rangle (1-f_{n_4})  + \langle n_3n_2 |
W^{>}(t, \bar t) |n_4n_5 \rangle f_{n_4} \Big) \nonumber \\
-\delta_{n_1n_3} & G^<_{n_4n_5}(\bar t)\Big( \langle n_3n_2 |
W^{<}(t, \bar t) |n_4n_5 \rangle f_{n_1} + \langle n_3n_2 | W^{>}(t, \bar t) |n_4n_5 \rangle
(1-f_{n_1}) \Big) \Big]. \nonumber
\end{align}

The equation for $\chi^{\mathrm{r}}$ is obtained by functional
differentiation of Eq.~\eqref{eqnmotG} with respect to $U(t')$ and
letting $U\to 0$ afterwards. This is done by replacing everywhere
the term $\delta G_{n_1n_2}^<(t)/\delta U_{n_3n_4}(t')$ with
$-i\hbar \langle n_1n_2 |\chi^{\mathrm r}(t-t')| n_3 n_4\rangle$. In
the correlation contributions, terms such as $\delta
W^{\gtrless}/\delta U$ are consistently neglected, because we assume
that the external potential is weak enough as not to cause changes
in the screening properties of the system. This is in agreement with
current developments in GW theory~\cite{LouieW0}.  The resulting
equation can be cast in the form
\begin{equation}\label{eqnmotchi_t}
( i \hbar \frac{\partial}{\partial t}
-\epsilon_{\alpha}^{\mathrm{HF}})  \chi^{
\mathrm{r}}_{\alpha\beta}(t,t') 
+ n_{\alpha}  \Big( \delta_{\alpha\beta}+ \sum_{\gamma}(v^{
\mathrm{dir}}_{\alpha\gamma} -v^{
\mathrm{exc}}_{\alpha\gamma})\chi^{ \mathrm{
r}}_{\gamma\beta}(t,t')\Big) 
= \sum_{\gamma}  \int_{-\infty}^{t} {\mathrm{ d}}\bar t
\Delta_{\alpha\gamma} (t,\bar t) \chi_{\gamma\beta}^{
\mathrm{r}}(\bar t, t'). 
\end{equation}
The correlation kernel $\Delta$ reads
\begin{align}\label{DELTA_T}
\Delta_{(n_1n_2) (n_3n_4)} (t, \bar t)= \int \frac{{\mathrm
d}\omega'}{2\pi} \Big[ - &e^{-\frac{i}{\hbar}(\hbar \omega '+ \tilde\epsilon_{n_1} -
\tilde\epsilon_{n_4}^*)(t-\bar t)} f_{n_1} \langle n_1n_2 |
W^{<}(\omega') |n_3n_4 \rangle\\ 
 -& e^{-\frac{i}{\hbar}(-\hbar \omega '+ \tilde\epsilon_{n_3} -
\tilde\epsilon_{n_2}^*)(t-\bar t)} f_{n_2} \langle n_1n_2 |
W^{<}(\omega') |n_3n_4 \rangle \nonumber \\
+\sum_{n_5} (1-f_{n_5})
\Big(  \delta_{n_1n_3} & e^{-\frac{i}{\hbar}(\hbar\omega '+
\tilde\epsilon_{n_3} - \tilde\epsilon_{n_5}^*)(t-\bar t)}  \langle
n_4n_2 | W^{<}(\omega') |n_5n_5 \rangle
\nonumber \\
+  \delta_{n_2n_4} & e^{-\frac{i}{\hbar}(-\hbar \omega '+
\tilde\epsilon_{n_5} - \tilde\epsilon_{n_4}^*)(t-\bar t)}  \langle
n_1n_3 | W^{<}(\omega') |n_5n_5 \rangle \Big) \Big]
\nonumber \\
+\Big[ <\longleftrightarrow > ,& f \longleftrightarrow
(1-f) \Big], \nonumber
\end{align}
where the last line indicates additional terms, in which  $W^<$ is
replaced by  $W^>$, $f$ by $1-f$ and $1-f$ by $f$.  The integral
over $\omega'$ in  Eq.~\eqref{DELTA_T} comes from the Fourier
transformation $W^{\gtrless}(t,\bar t)=\int \frac{{\mathrm
d}\omega'}{2\pi}e^{-i\omega'(t-\bar t)}W^{\gtrless}(\omega')$ of the
screened  potential.

The particular time-dependence of the correlation contributions in
Eq.~\eqref{eqnmotchi_t} allows us to perform the Fourier
transformation with respect to time and to determine the
frequency-dependent   $\chi^{\mathrm r}$ by
\begin{equation}\label{eqnmotchi}
( \hbar  \omega-\epsilon_{\alpha}^{\mathrm{HF}})\chi^{
\mathrm{r}}_{\alpha\beta}(\omega) + n_{\alpha}\Big(
\delta_{\alpha\beta}+\sum_{\gamma}(v^{ \mathrm{dir}}_{\alpha\gamma}
-v^{ \mathrm{exc}}_{\alpha\gamma})\chi^{ \mathrm{
r}}_{\gamma\beta}(\omega)\Big)
= \sum_{\gamma} \Delta_{\alpha\gamma} (\omega) \chi_{\gamma\beta}^{
\mathrm{r}}(\omega).
\end{equation}
Using  
\begin{equation}
\int_{-\infty}^{t} {\mathrm{ d}}\bar t
e^{-\frac{i}{\hbar}(\omega\pm \epsilon+i\gamma)\bar t}= \frac{i\hbar
e^{-\frac{i}{\hbar}(\omega\pm\epsilon+i\gamma)t}}{\omega\pm \epsilon
+ i\gamma},
\end{equation}
 integration over $\bar t$ yields for the correlation
kernel
\begin{align}\label{firstD}
&\qquad \Delta_{(n_1n_2)(n_3n_4)}(\omega) = i\hbar \int
\frac{\mathrm{d}\omega'}{2\pi}\times \\
&\quad \Big(  \frac{ f_{n_1} }{\hbar\omega -\hbar\omega' -\tilde
\epsilon_{n_1}+\tilde \epsilon^*_{n_4}}\langle n_1n_2 | W^<(\omega')
|n_3n_4 \rangle \nonumber \\
&\qquad +  \frac{f_{n_2}  }{\hbar\omega +\hbar\omega' - \tilde
\epsilon_{n_3}+ \tilde \epsilon^*_{n_2}}\langle n_1n_2 |
W^<(\omega') |n_3n_4
\rangle \nonumber\\
&\quad - \sum_{n_5}  \frac{(1-f_{n_5})  \delta_{n_1n_3}
}{\hbar\omega -\hbar\omega' -\tilde \epsilon_{n_3}+\tilde
\epsilon^*_{n_5}}\langle n_4n_2 | W^<(\omega') |n_5 n_5
\rangle\nonumber \\
&\qquad - \sum_{n_5}  \frac{(1-f_{n_5})  \delta_{n_4n_2}
}{\hbar\omega +\hbar\omega' -\tilde \epsilon_{n_5}+\tilde
\epsilon^*_{n_4}}\langle n_1n_3 | W^<(\omega') |n_5n_5 \rangle \Big)
\nonumber \\
& \qquad \qquad +\Big[ \big(<\rightarrow > \big), \big(f
\leftrightarrow (1-f)\big) \Big]. \nonumber
\end{align}
Further, the functions $W^{\gtrless}$ are related to Im$W^{\mathrm
r}$ through the Kubo-Martin-Schwinger boundary
conditions~\eqref{KMS_W}. Finally, the retarded screened potential
needs to be determined from the density response function via the
discrete version of Eq.~\eqref{Weps}, i.e., by using
\begin{align}\label{W:chiandW}
&\langle n_1 n_2 |W^{\mathrm{r}}(\omega)|n_3 n_4 \rangle =\langle
n_1n_2 |v|n_3n_4 \rangle \\
& \quad  + \sum_{n_5\dots n_8} \langle n_1n_5 |v|n_3n_7\rangle
\langle n_5n_6|\chi^{\mathrm{r}}(\omega)|n_7n_8\rangle \langle
n_2n_6|v|n_4n_8 \rangle.\nonumber
\end{align}

Here, we use for $\chi^{\mathrm{r}}$ the Lindhard polarization
\begin{equation}\label{Lindhard}
\chi_{\alpha\beta}^{ 0}(\omega)= - \delta_{\alpha\beta}\frac{
n_{\alpha}}{(\hbar\omega-\epsilon^{\mathrm{HF}}_{\alpha} +
i\Gamma)},
\end{equation}
with $\Gamma \to 0$. This yields a $\delta$ distribution function
for $\mathrm{Im}\chi^0$ and allows one to analytically evaluate the
$\omega'$ integral in Eq.~\eqref{firstD}. This choice for
$\chi^{\mathrm{r}}$ is consistent with our earlier assumption in the
derivation of Eq.~\eqref{eqnmotchi} that the screening properties
are unchanged by the external potential.

Equations~\eqref{eqnmotchi} and~\eqref{firstD} complete the
development of our method: Together with Eq.~\eqref{KMS_W}, they
determine $\chi^{\mathrm r}$ and therefore via Eq.~\eqref{Weps} the
retarded screened potential $W^{\mathrm r}$. This in turn enters the
calculation of the equilibrium Green's function via
Eqs.~\eqref{Dyson} and~\eqref{ImSigma}. In Eq.~\eqref{firstD}, the
real part of $\Delta$ contributes to transition-energy
renormalizations and the imaginary part to resonance broadening. The
diagonal contributions $\Delta_{\alpha=\gamma}$ only shift and
broaden two-particle resonances whereas the off-diagonal
$\Delta_{\alpha\neq\gamma}$ together with
$v^{\mathrm{dir}}_{\alpha\gamma}$ and
$v^{\mathrm{exc}}_{\alpha\gamma}$ can lead to collective features in
the $\mathrm{Im}\varepsilon^{-1}(\omega)$ spectrum. We reiterate
that the quasiparticle properties are conserving on the one and
two-particle levels in the sense of Baym and Kadanoff because the
one-particle Green's function is calculated from a dielectric
function (density response-function) that is related by a functional
differentiation to a one-particle conserving equation of motion for
$G^<$. The finite damping of the resonances in the
$\varepsilon^{-1}$ spectrum results not only from the broadening of
electronic quasiparticle states but also from the inclusion of
correlation effects in the equation for $\chi^{\mathrm r}$.

\subsection{Relation to other methods}

Equation~\eqref{eqnmotchi} has a structure reminiscent of a
Bethe-Salpeter (BS) equation. The similarities and differences
between the present approach and GW-based BS calculations are
discussed in Appendix B.

From Eq.~\eqref{eqnmotchi}, one can determine the density-density
correlation function and thus $\varepsilon^{-1}$ in different
approximations. First, neglecting all Coulomb contributions and
setting $\Delta_{\alpha\gamma}(\omega)$ $ \to i
\delta_{\alpha\gamma} \Gamma$, one obtains the Lindhard polarization
$ \chi^{ 0}$ given in Eq.~\eqref{Lindhard}. Second, if only
$v^{\mathrm{dir}}$ in the Coulomb enhancement contribution is
included, then Eq.~\eqref{eqnmotchi} takes the form of a BS equation
in the ladder approximation~\cite{Baym-Kadanoff:PR,Kwong-Bonitz}.
Using Eq.~\eqref{Weps} with $\chi^{\mathrm r}$  computed at that
level corresponds to the RPA for $W^{\mathrm{r}}$ with HF Green's
functions in the irreducible polarization function, i.e.,
\begin{align}
\label{W-RPA} W^{\mathrm{r}}=v+ vP^{\mathrm{RPA}}
\,W^{\mathrm{r}}\\
P^{\mathrm{RPA}}= -i\hbar G^{\mathrm{HF}} G^{\mathrm{HF}}. 
\end{align}
We will refer  to this approximation in the following as GW-RPA.

Including $v^{\mathrm{exc}}$ in addition to $v^{\mathrm{dir}}$ in
the Coulomb enhancement term~\eqref{Rabi} means taking into account
mean-field exchange effects on the density response. The additional
inclusion of the scattering kernel $\Delta(\omega)$ incorporates
correlations beyond the mean-field level. We will call this
calculation procedure ``consistent GW'' in the following, because
the screened potential is calculated using an approximation that
corresponds to the GW approximation for the self-energy
$\Sigma^{\mathrm r}$ (see Eq.~\eqref{GW_time}).

\section{Numerical results}\label{Section3}

We discuss the characteristics of the GW-RPA and the consistent GW
method using numerical results for small sodium clusters. The
eigenfunctions and eigenvalues of the ground state are calculated by
performing a stationary self-con\-sis\-tent field calculation for a
closed-shell configuration~\cite{YaroslavHubner}. For the Na atoms
we use the \textsc{lanl2dz} basis set (Dunning-Huzinaga full double
zeta on the first row, Los Alamos effective core potential plus
double zeta on Na--Bi)~\cite{LosAlamos}, extended to improve the
description of low-lying states above the Fermi level. Thus, one
valence electron of the Na atoms is represented by 15 basis
functions in the (6s/3p) configuration. Since we are mainly
interested in the energy range around the Fermi energy, the deeply
lying states are incorporated in the effective core potential for
the inner electrons. In this way, each Na atom contributes with two
electrons with opposite spin and we treat 2, 4, and 10 doubly
occupied molecular orbitals for Na$_4$, Na$^+_9$, and Na$_{21}^+$,
respectively. For the calculation of the screened Coulomb potential
and the GW self-energy, we use 50 HF spin orbitals.

We will first present results for Na$^+_9$, where the first four
spin orbitals are doubly occupied in the ground state and with our
choice of the effective core potential. The discrete HF energy
spectrum contains a gap between highest occupied (HOMO) and the
lowest unoccupied molecular HF orbital (LUMO) of
$\epsilon^{\mathrm{HF}}_{\mathrm{LUMO}}-\epsilon^{\mathrm{HF}}_{\mathrm{
HOMO}}=4.497$\,eV, which is large compared to the level-spacing.

\begin{figure}
\resizebox{0.48\textwidth}{!}{\includegraphics{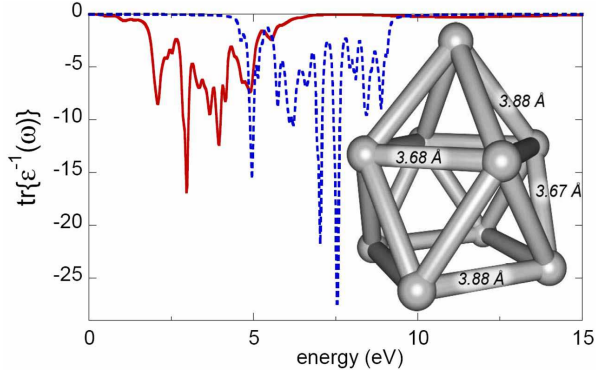}}
\caption{Trace over the  inverse dielectric function  in the  GW-RPA
(dashed line), and consistent GW (solid line) approaches for
Na$_9^+$. The inset shows the geometric structure used.}
\label{fig:IMe}
\end{figure}

Before considering the QP properties of electrons, we examine the
properties of the screened potential $W^{\mathrm r}(\omega)$.
Figure~\ref{fig:IMe} shows the trace over the imaginary part of the
inverse dielectric function
\begin{equation}
\sum_{\alpha}{
\mathrm{Im}}\,\varepsilon^{-1}_{\alpha\alpha}=\sum_{\alpha} (v\cdot
{ \mathrm{Im}}\chi^{\mathrm{ r}})_{\alpha\alpha}.
\end{equation}
In the GW-RPA method, the finite width of the peaks for
$\varepsilon^{-1}$ is due to a phenomenological damping of $\Gamma =
0.1~\mathrm{eV}$ instead of the scattering term in
Eq.~\eqref{eqnmotchi}. Several peaks arising from resonances in the
discrete level system are visible in the spectrum. We compare the
RPA result with the consistent GW case, calculated using in
Eq.~\eqref{firstD} for $\Delta$ a quasiparticle broadening of
$\gamma=0.25$\,eV for the HF energies.  One observes a drastic red
shift of the whole spectrum by about 3 eV, together with a
redistribution of spectral weight and a decrease of the
``bandwidth'' of the imaginary part of the inverse dielectric
function, a typical correlation effect. This trend is in agreement
with Ref~\cite{ShirleyMartin}, where a sizable change of QP energies
in atomic systems was obtained when going from the GW-RPA to a GW
calculation including exchange effects in the density-density
correlation function.

\begin{figure}
\resizebox{0.6\textwidth}{!}{\includegraphics{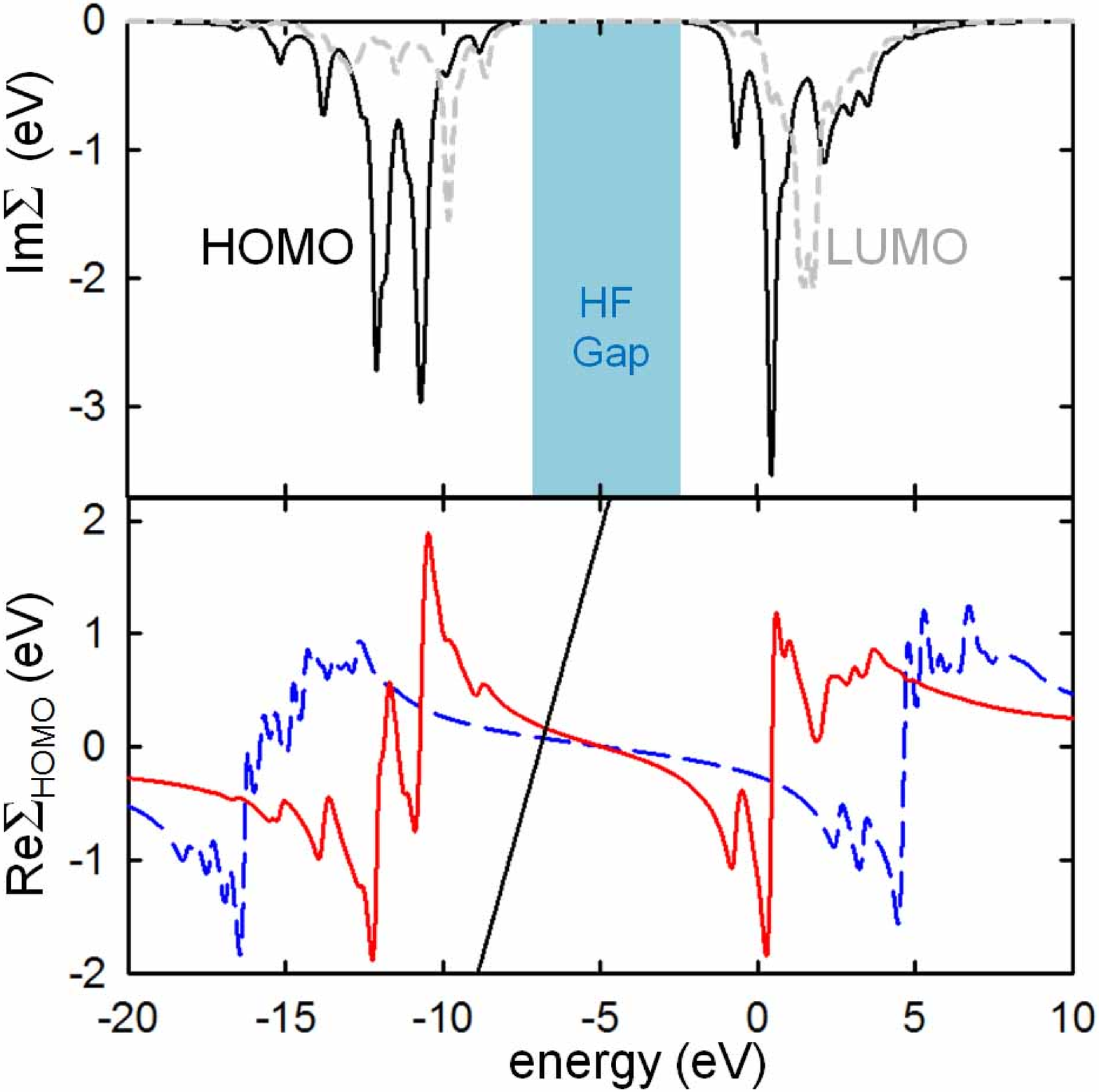}}
\caption{Upper panel: Imaginary part of the self-energy for the HOMO
(solid black line) and the LUMO (dashed grey line) of Na$_9^+$,
calculated in consistent GW approximation. Lower panel: The real
part of the self-energy for the HOMO in GW-RPA (dashed blue line)
and consistent GW (solid red line).} \label{fig:HO-LU}
\end{figure}

That this change is an  improvement of the dynamically screened
potential in the consistent GW calculation over the GW-RPA result is
substantiated by evaluating the \emph{f}-sum rule for finite
systems. This relation  is a measure of the quality of an
approximation for $\varepsilon^{-1}$ and is directly related to the
particle number conservation law at the two-particle level. A
derivation of the \emph{f}-sum rule for finite systems  is presented
in Appendix~\ref{AppendixA}. We find that for the consistent GW
calculation, the $f$-sum rule Eq.~\eqref{f:fsumFINAL} is fulfilled
to better than $0.05\%$, whereas the GW-RPA violates it by 280\%.
This is a clear indication of the well-known fact that the RPA
approximation is consistent \emph{only} with Hartree Green's
functions~\cite{Baym-Kadanoff:PR,TammeSchepe} and inconsistent for
other approximate single-particle Green's
functions~\cite{Szymansky}. This restriction and inconsistency is
removed at the consistent GW level.

Next, we use $W^{\mathrm r}$ on the level of GW-RPA and consistent
GW, respectively, as input to the  GW calculation of the QP
properties. As a numerical check for the computational procedure,
the normalization of the spectral function, i.e.,  particle number
conservation on the single-particle level, is fulfilled to better
than 0.1\%.

\begin{figure}
\resizebox{\textwidth}{!}{\includegraphics{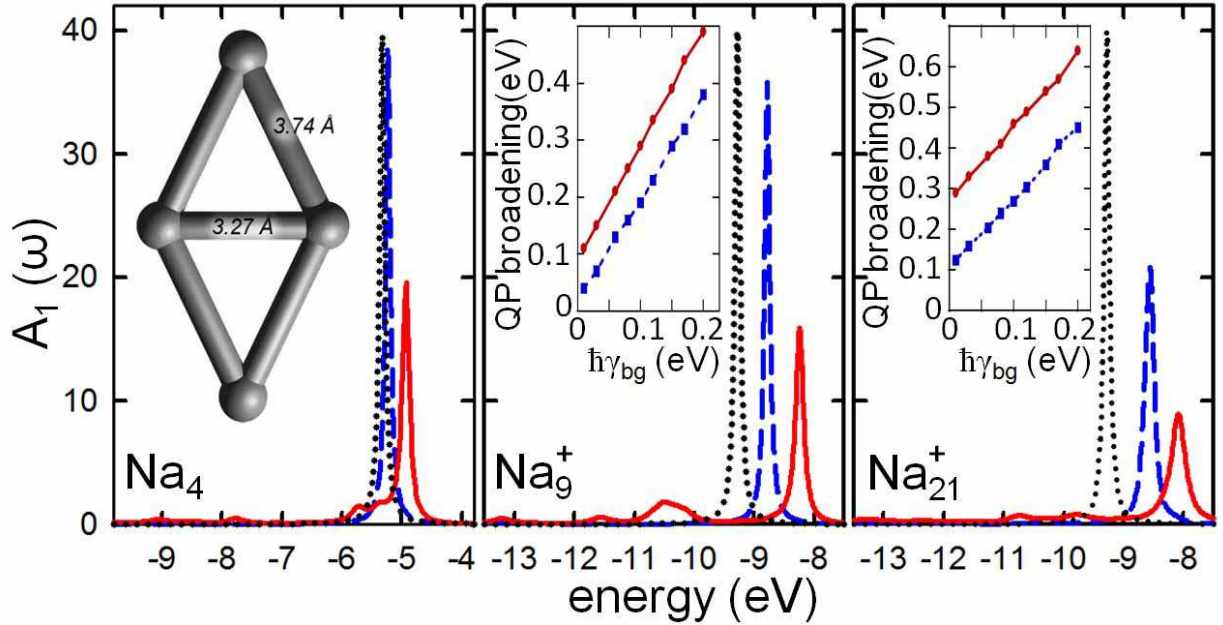}}
\caption{Spectral function of the single-particle state with lowest
energy for Na$_4$, Na$_9^{+}$ and Na$_{21}^{+}$ for HF (dotted black
line), GW-RPA (dashed blue line) and consistent GW (solid red line)
calculated with $\hbar \gamma_{\mathrm{bg}}=0.05$~eV. The insets
show the geometric structure used for Na$_4$, and the dependence of
the QP broadening on the background damping for Na$_9^{+}$ and
Na$_{21}^{+}$.} \label{fig:specfunc}
\end{figure}

Concerning the QP properties, we first discuss properties of the
HOMO and LUMO in Na$_9^+$. The upper panel of Fig.~\ref{fig:HO-LU}
depicts the energy dependence of $\mathrm{Im} \Sigma^{\mathrm{r}}$
in the consistent GW approximation. Inside the gap, it drops nearly
to zero, leading to long lifetimes for states around the HOMO-LUMO
gap (in agreement with the results from Ref~\cite{YaroslavHubner}).
The renormalized single-particle energies are obtained by solving
the Dyson equation Eq.~\eqref{Dyson} where we add a constant
background damping $i\hbar\gamma_{\mathrm{bg}}$ to $\Sigma^{\mathrm
{corr}}(\omega)$ to avoid numerical problems due to the small
damping near the HOMO-LUMO gap. The lower panel of
Fig.~\ref{fig:HO-LU} shows the real part of the retarded self-energy
$\Sigma^{\mathrm{corr}}$. To determine the QP energies, we note that
QP poles in the retarded Green's function in Eq.~\eqref{Dyson}
result if the real part of the denominator vanishes, i.e., if
\begin{equation}\label{geometry}
\hbar\omega-\epsilon_n^{\mathrm{HF}}
=\Sigma^{\mathrm{corr}}_{nn}(\omega).
\end{equation}
The solution of Eq.~\eqref{geometry} corresponds to finding the
intersection of the function
$\mathrm{Re}\Sigma^{\mathrm{corr}}(\omega)$ with the straight line
$\hbar\omega-\epsilon^{\mathrm{HF}}_n$. This line is also shown in
Fig.~\ref{fig:HO-LU}.

The QP broadening is then the value of
$\mathrm{Im}\Sigma^{\mathrm{corr}}$ at the QP energy. The GW-RPA and
consistent GW lead to different predictions for the size of the
energy difference between the  HOMO and the LUMO. In
Table~\ref{tableGAP} we compare the calculated values of the gap
with other theories and also experiment for the cases when data  is
available in the literature. Notice that in the case of Na$_4$ our
method yields a  HOMO-LUMO gap of 3.4\,eV which is in very good
agreement with the experimental value of 3.35\,eV  from
photoelectron spectroscopy~\cite{exp1,exp2}.


\begin{table}
\caption{The values of the energy gaps (in eV) between the HOMO and
the LUMO for the studied Na clusters.} \label{tableGAP}
\begin{tabular}{c|ccc}
\hline \hline
& Na$_4$ &  Na$_9^+$ & Na$_{21}^+$ \\
\hline
HF & 3.60 & 4.49 & 2.71 \\
GW-RPA & 3.56 & 4.31 & 2.59 \\
consistent GW & 3.40 & 4.14 & 2.44 \\
\hline
literature & 3.00\,$^a$ & 3.38\,$^b$ &  \\
& 3.35$\pm$0.2\,$^c$ &  &  \\
\hline \hline  \multicolumn{4}{l} {$^a$\,DFT+GW-RPA, from Ref.~\cite{onida-prl}.} \\
\multicolumn{4}{l} {$^b$\,self-consistent GW-RPA, from Ref.~\cite{YaroslavHubner}.} \\
\multicolumn{4}{l} {$^c$\,experiment, from Ref.~\cite{exp1,exp2}.}
\end{tabular}
\end{table}

From the spectral function, one can read off level shifts,
broadening and  redistribution of spectral weight. These features,
important for states away from the HOMO and the LUMO, are
experimentally accessible. Figure~\ref{fig:specfunc} shows the
spectral functions of the single-particle state with the lowest
energy for  Na$_4$, Na$_9^+$ and Na$_{21}^+$, i.e., HOMO$-1$,
HOMO$-3$ and HOMO$-9$, respectively. The insets show the dependence
of the width (FWHM) of the QP peak on $\gamma_{\rm bg}$ used in the
self-consistent GW calculation. Extrapolating to zero, one obtains
the QP widths due to Coulomb correlations: for Na$_9^+$, the QP
broadening is 0.085 (0.02) eV for the consistent GW (GW-RPA). For
Na$_{21}^+$, the  QP broadening is 0.29 (0.11) eV for the consistent
GW (GW-RPA). For Na$_4$, this intrinsic QP width is on the order of
the background broadening $\hbar\gamma_{\mathrm{bg}} = 0.03$\,eV and
an extrapolation would be less accurate. While GW-RPA shows only a
weak broadening of the QP peak, the consistent GW, as expected,
yields a much broader main QP peak, together with a more pronounced
redistribution of spectral weight that reaches lower energies with
increasing cluster size. In terms of lifetimes, the QP broadenings
of the lowest energy states give 7.74~fs for Na$_9^+$ and 2.26~fs
for Na$_{21}^+$, in the case of the consistent GW approach. The
order of magnitude of the lifetime values is in agreement with
experiment, where the lifetime of the plasmon resonance in
Na$_{93}^+$ was found to be 10~fs;~\cite{expLifetime1} also, in
second harmonic generation time-resolved measurements on larger
surface-supported Na clusters the same lifetime was obtained for a
cluster size of about 25~nm~\cite{SGHlifetime}.

\section{Conclusions}

In conclusion, we have studied the effect of Coulomb correlations on
the QP properties of electrons in metallic clusters within the
framework of GW theory. Employing a linear-response calculation of
the density-density correlation function allows us to obtain a
conserving dynamically screened potential including finite,
non-phenomenological damping. We analyze two approximations for the
kinetic equation for the density response. One corresponds to the
standard GW-RPA for the dielectric function based on HF energies and
wave functions. The other one includes mean-field direct and
exchange as well as correlation contributions to the dielectric
function and the self-energy. The latter approximation is based on a
consistent treatment of one- and two-particle correlations within
the GW approximation. It is conserving in the sense of Baym and
Kadanoff on the one and two-particle levels, and therefore fulfills
the $f$-sum rule. Compared to GW-RPA, we find differences in the
spectral function peak positions of up to 1\,eV and differences in
the QP broadening of more than a factor of 3 due to modifications of
the screened Coulomb potential.

G.P and W.H. would like to acknowledge the support from the
Schwerpunktprogramm SPP 1153 of the German Research Foundation.

\appendix

\section{The $f$-sum rule for finite systems}\label{AppendixA}

In this Appendix we show how the $f$-sum rule for the
density-density correlation function or, equivalently, for the
inverse dielectric function for finite systems follows from
particle-number conservation. To this end, we define the quantity
\begin{equation}\label{A:defineC}
C(\vec r_1,\vec r_2;t_1)\equiv i\hbar \frac{\partial}{\partial
t_1}\left< [ \rho(1), \rho(2)] \right>_{t_1=t_2},
\end{equation}
where the angular brackets indicate an equilibrium averaging, and
$\rho(1)=\psi^{\dag}(1)\psi(1)$ is the particle density operator
expressed by the field creation and destruction operators. We then
evaluate Eq.~\eqref{A:defineC} using the particle-number
conservation condition, and finally relate $C$ to the Fourier
transform of the density-correlation function.

To evaluate Eq.~\eqref{A:defineC}, we employ the continuity equation
\begin{equation}\label{A:continuity}
\frac{\partial  \rho(1) }{\partial t_1}+ \nabla_1 \cdot  \vec \jmath
(1) = 0,
\end{equation}
which is a statement of particle-density conservation in operator
form. Here, the current density operator is given by
\begin{equation}\label{A:current}
\vec \jmath(1)=\left. \frac{\hbar}{2im} \left( \nabla_1
-\nabla_{1'}\right) \psi^{\dag}(1')\psi(1) \right|_{1'=1^+}.
\end{equation}
Using~\eqref{A:continuity} and~\eqref{A:current},
Eq.~\eqref{A:defineC} yields
\begin{equation}\label{A:beginC}
\qquad C(\vec r_1,\vec r_2;t_1) = \\
 \frac{\hbar^2 }{2m} \left<  \big[ \nabla_1^2 \psi^{\dag}(1)\psi(1)
- \psi^{\dag}(1) \nabla_1^2 \psi(1), \rho(2)\big] \right>_{t_1=t_2}.
\end{equation}
Computing the commutator in a straightforward manner, one obtains
\begin{align}\label{A:Cbegindelta}
C (\vec r_1,\vec r_2;t_1) = & \frac{\hbar^2}{2m}\big[ \left<  \nabla^2_1 \psi^{\dag}(1) \psi(2)
+ \psi^{\dag}(2) \nabla^2_1 \psi(1)  \right>_{t_1=t_2}
\delta(\vec r_1-\vec r_2) \\
& \quad - \left<  \psi^{\dag}(1) \psi(2) + \psi^{\dag}(2) \psi(1)
\right>_{t_1=t_2} \nabla^2_1 \delta(\vec r_1-\vec r_2) \big].
\nonumber
\end{align}
This expression cannot be treated further without integrating over
$\vec r_1$ and $\vec r_2$ with smooth but otherwise arbitrary test
functions. Choosing the  test functions to be plane waves $e^{-i
\vec q \cdot \vec r_1}$ and $e^{ i \vec q \cdot \vec r_2}$, we
evaluate the quantity
\begin{equation}\label{A:defCqq}
C_{\vec q \vec q}(t_1) =\int  e^{-i \vec q \cdot \vec r_1} C (\vec
r_1,\vec r_2;t_1)  e^{ i \vec q \cdot \vec r_2} {\mathrm d}^3 r_1
{\mathrm d}^3 r_2.
\end{equation}
From Eq.~\eqref{A:Cbegindelta}, one has
\begin{align}
C_{\vec q \vec q}  (t_1) = \mbox{}&\frac{\hbar^2}{2m} \int  \left<
\nabla^2_1 \psi^{\dag}(1) \psi(1) + \psi^{\dag}(1) \nabla^2_1
\psi(1) \right> {\mathrm d}^3 r_1
\nonumber \\
& \mbox{}-   \frac{\hbar^2}{2m} \int \nabla^2_1 \big[ e^{-i \vec q \cdot
\vec r_1} \left< \psi^{\dag}(1)
\psi(2) + \psi^{\dag}(2) \psi(1) \right>_{t_1=t_2}  \big] \delta(\vec r_1-\vec r_2)   e^{ i \vec q \cdot \vec
r_2} {\mathrm d}^3 r_1 {\mathrm d}^3 r_2,
\end{align}
which yields
\begin{equation}
C_{\vec q \vec q}  (t_1) =- \frac{\hbar^2}{m}  \int \big[ q^2 \left<
\rho(1) \right>  -i\vec q \cdot \nabla_1 \left< \rho(1) \right>
\big]{\mathrm d}^3 r_1.
\end{equation}
This integral effectively extends over a finite volume because the
integrand $\rho$ is the charge density of a finite system. The
second contribution to the integral vanishes because it can be
transformed into an integral over a closed surface lying outside of
the charge density. The first integral is straightforward and yields
\begin{equation}\label{A:Cqqend}
C_{\vec q \vec q}(t_1) = \frac{\hbar^2 q^2}{m} N_{\mathrm e},
\end{equation}
where $N_{\mathrm e}$ is the total number of electrons in the
system.

We now need to relate the commutator $C$ to the den\-si\-ty-density
correlation function, which is defined as
\begin{equation}\label{A:def_chi}
i\hbar \chi^{\mathrm r} (1,2) = \Theta(t_1-t_2) \langle \left[
\rho(1), \rho(2) \right] \rangle.
\end{equation}
In equilibrium, the correlation function depends only on the
difference of the times $t_1$ and $t_2$~\cite{Kadanoff-Baym:Book}.
Using this property, one performs a Fourier transformation in
$t_1-t_2$ and obtains for the imaginary part of $ \chi^{\mathrm
r}(\omega)$
\begin{equation}\label{A:ImchiFT}
{\mathrm{ Im}}  \chi^{\mathrm r} (\vec r_1,\vec r_2;\omega)=
-\frac{1}{2\hbar} \langle \left[ \rho(\vec r_1,\omega), \rho(\vec
r_2) \right] \rangle .
\end{equation}
On the other hand, Fourier transformation of Eq.~\eqref{A:defineC}
yields
\begin{equation}
C(\vec r_1,\vec r_2;t_1)= i\hbar\int_{-\infty}^{+\infty}
\frac{\mathrm d \omega}{2 \pi} \omega \, \langle \left[ \rho(\vec
r_1,\omega), \rho(\vec r_2) \right] \rangle ,
\end{equation}
so that we obtain
\begin{equation}\label{A:relateC2Chi}
C(\vec r_1,\vec r_2;t_1) = -\frac{\hbar^2}{\pi}
\int_{-\infty}^{+\infty}  {\mathrm d}\omega \, \omega {\mathrm{ Im}}
\chi^{\mathrm r} (\vec r_1, \vec r_2;\omega).
\end{equation}
Inserting this result for $C$ in~\eqref{A:defCqq} and equating it
with~\eqref{A:Cqqend}, one obtains
\begin{equation}\label{fsumrule1}
\frac{q^2}{m} N_{\mathrm e} = - \frac{1}{\pi}
\int_{-\infty}^{+\infty} 
{\mathrm d}\omega\omega
\int {\rm e}^{-i \vec q\cdot \vec r_1} {\rm Im} \chi^{\rm r}
(\vec r_1,\vec r_2;\omega) {\rm e}^{i \vec q \cdot \vec r_2}
{\mathrm d}^3 r_1{\mathrm d}^3 r_2 .
\end{equation}
The $f$-sum rule for finite systems can now be derived by expanding
$\chi^{\mathrm{r}}$ on the RHS of Eq.~\eqref{fsumrule1} into the
basis functions (HF orbitals) $\{ \varphi_n(\vec r) \}$. Thus
\begin{equation}\label{f:rule1}
- \frac{2}{\pi} \int_{0}^{\infty} d\omega \, \omega  \sum_{n_1\dots
n_4}
P_{n_1n_3}(-\vec q)  P_{n_2n_4}(\vec q)
 \mathrm{Im}\left< n_1n_2 \left| \chi^{\rm r}(\omega) \right|
n_3n_4 \right> =\frac{q^2}{m}N_{\rm e},
\end{equation}
where we have defined the overlaps
\begin{eqnarray}
P_{n_1n_2}(\vec q)=\int {\rm e}^{ i\vec q\cdot \vec r}
\varphi^*_{n_1}(\vec r) \varphi_{n_2}(\vec r) {\mathrm d}^3 r
\end{eqnarray}
and used the symmetry of
$\chi^{\mathrm{r}}(-\omega)=\chi^{\mathrm{r}}(\omega)^{*}$.
Expanding the exponential function yields
\begin{equation}\label{f:Pexpansion}
P_{n_1n_2}(\vec q) = \int \varphi^*_{n_1}(\vec r) \left( i \vec q
\cdot \vec r \right)
\varphi_{n_2}(\vec r)\, {\mathrm d}^3r + 
\text{terms of higher order in }q .
\end{equation}
Note that there is no $q$-independent term because the basis
functions $\{  \varphi_n (\vec r) \}$ are orthogonal for $n_1 \neq
n_2$. Using Eq.~\eqref{f:Pexpansion} in Eq.~\eqref{f:rule1},
dividing by $q^2$ and then letting $q\rightarrow 0$ yields the
$f$-sum rule for finite systems:
\begin{equation}\label{f:fsumFINAL}
\int_{0}^{\infty}  {\rm d} \omega \, \omega  \sum_{n_1\dots n_4}
\left( \vec n \cdot \vec r_{n_1n_3} \right) \left(
\vec n \cdot \vec r_{n_2n_4} \right)
\left[- \mathrm{Im} \langle n_1n_2|\chi^{\rm
r}(\omega)|n_3n_4 \rangle \right]= \frac{\pi}{2 m}N_e .
\end{equation}
Here $\vec n$ is a unit vector in an arbitrary direction and we have
defined the matrix element of the position operator
\begin{equation}
\vec r_{n_1n_2}=\int \varphi^*_{n_1}(\vec r)\, \vec r \,
\varphi_{n_2}(\vec r) \, {\mathrm d}^3 r .
\end{equation}

\section{Comparison with Bethe-Salpeter Equation} \label{AppendixB}

In this Appendix, we wish point out the differences and similarities
of our calculation to the BS equation approach, which has been
employed for the calculation of absorption spectra in a variety of
systems, including finite systems such as Na$_4$~\cite{onida-prl}.

The following derivations  follow closely the arguments of
Ref.~\cite{Baym-Kadanoff:PR}, which uses equilibrium one- and
two-particle Green's functions for complex times. Consequently, the
time integrations extend over $0<it<1/(k_{\mathrm{B}}T)$, where
$k_{\mathrm{B}}$ is Boltzmann's constant and $T$ is the
temperature~\cite{Kadanoff-Baym:Book}. To extract the real-time
Green's functions from this formalism one needs to perform an
analytical continuation to real times. An equivalent procedure,
which is closer to the formal development in this paper, is to keep
the notation of Ref.~\cite{Baym-Kadanoff:PR}, but to interpret the
Green's functions as real-time Green's functions defined on the
Keldysh contour, and the integrals over time as Keldysh contour
integrals.

We work with correlation functions that are dependent on space-time
variables instead of the the quantities labeled by HF spin orbitals
introduced in Eqs.~\eqref{Gnn} and~\eqref{indexCoulomb}. We begin by
defining the four-point correlation function
\begin{equation}
i\hbar L(12,34) = \langle T [\psi(1)\psi(3)
                              \psi^{\dag}(2)\psi^{\dag}(4)]\rangle 
-   \langle T [\psi(1)\psi^{\dag} (2) ] \rangle \langle T
[\psi(3)\psi^{\dag} (4) ]\rangle,
\end{equation}
where $T$ ensures the time-ordering on the Keldysh contour.
Introducing an external disturbance $U(3,4)$ one can generate the
four-point function by the functional derivative
\begin{equation}\label{B:L}
L(12,34) = -i\hbar\left. \frac{ \delta G (1,2)}{\delta
U(3,4)}\right|_{U=0}.
\end{equation}
The density-density correlation function used earlier in this paper
is given in terms of the four-point correlation function by
\begin{equation}
\label{chi-L} \chi(1,2)=L(11^+,22^+)
\end{equation}
where $1^+ = (\vec{r}_1\sigma_1t_1^+)$ and $t_1^+$ is a time on the
Keldysh contour infinitesimally larger than $t_1$. The screened
potential $W$ is then determined by
\begin{equation}
\label{Wv-chi-v}
 W(1,2)= v(1,2)+ V(1,\bar{3})\chi(\bar{3},\bar{4})V(\bar{4},2).
\end{equation}
Here and in the following the convention is used that repeated
space, spin, and time indices with an overbar are integrated over
space and time and summed over spin indices. Further, the
abbreviation
\begin{equation}
V(1,2) = \delta(t_1 - t_2)
\delta_{\sigma_1,\sigma_2}v(\vec{r}_1-\vec{r}_2)
\end{equation}
has been introduced. For the case of the independent-particle
approximation, this correlation function is given by
\begin{equation}\label{B:L0}
L^0(12,34) = -i\hbar G(1,3)G(4,2).
\end{equation}

\begin{figure}
\resizebox{0.6\textwidth}{!}{\includegraphics{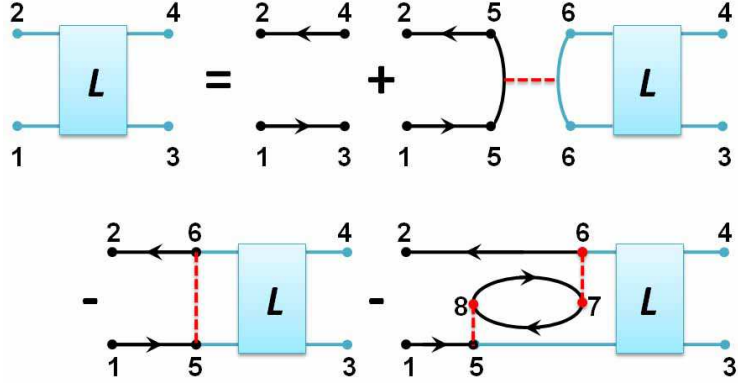}}
\caption{Diagrammatic representation of the BS
  equation~\eqref{L-K}. The square represents the four-point
  correlation function defined in Eq.~\eqref{B:L}, the two
  electron-hole lines stand for the independent particle correlation
  function, and the dashed lines are for the bare Coulomb potential.}
\label{fig:BSE}
\end{figure}

To arrive at an equation for $L$, one starts with the Dyson equation
for the Green's function:
\begin{equation}\label{B:Dyson}
G^{-1}(1,2)=G^{-1}_{0}(1,2)-U(1,2)-\Sigma(1,2).
\end{equation}
In the GW approximation, the self-energy  consists of the unscreened
direct (Hartree) term and the screened exchange (GW) term:
\begin{equation}\label{B:Sigma}
\Sigma(1,2)=-i\hbar  \delta(1-2) V(1,\bar{3})G(\bar{3},\bar{3}^{+})
+  i\hbar G(1,2)W(2,1)
\end{equation}
From the identity  $G(1,\bar{3})G^{-1}(\bar{3},2)=\delta(1-2)$, it
follows
\begin{equation}\label{B:identity}
\frac{\delta G(1,2)}{\delta U(3,4)}= -G(1,\bar{5}) \frac{\delta
G^{-1}(\bar{5},\bar{6})}{\delta U(3,4)} G(\bar{6},2).
\end{equation}
After calculating the functional derivative of the inverse Green's
function with respect to $U$ from Eq.~\eqref{B:Dyson}, one obtains
\begin{equation}\label{B:dGdU1}
\frac{ \delta G (1,2)}{\delta U(3,4)} = \mbox{}  G(1,\bar{5})
G(\bar{6},2)  \Big[
  \delta(3-\bar{5})
\delta(4-\bar{6}) + \frac{\delta \Sigma(\bar{5},\bar{6})}{\delta
U(3,4)} \Big].
\end{equation}
Inserting in Eq.~\eqref{B:dGdU1} the self-energy
from~\eqref{B:Sigma}, one obtains after performing the derivative
with respect to $U$ (neglecting terms with $\delta W/\delta U$) and
using the definition~\eqref{B:L}, an equation for the four-point
correlation function function that is consistent with the GW
self-energy
\begin{equation} \label{L-K} 
L(12,34)= L^0(12,34)
+ L^0(12,\bar{5} \bar{6}) 
 [K_{\mathrm{HF}}(\bar{5}\bar{6},\bar{7}\bar{8})+
K_{\mathrm{c}}(\bar{5}\bar{6},\bar{7}\bar{8})]L(\bar{7}\bar{8},34)
\end{equation}
In Eq.~(\ref{L-K}), the contributions to the kernel consist of the
\emph{bare} Hartree-Fock contribution
\begin{equation}
  K_{\mathrm{HF}}(12,34) = \delta(1-2)\delta(3-4) V(1,3)
  - \delta(1-3)\delta(2-4)V(1,2),
\label{K-HF}
\end{equation}
and the correlation term
\begin{equation}
K_{\mathrm{c}}(12,34) =  -  \delta(1-3)\delta(2-4)V(1,\bar{5})
\chi(\bar{5},\bar{6}) V(\bar{6},2). \label{Kc1}
\end{equation}
It is apparent that the screened Coulomb interaction enters only in
the correlation contribution, which can be evaluated after the
density-density correlation function is specified. A natural
starting point is the noninteracting density-density correlation
function
\begin{equation}
\chi^0(1,2)=-i \hbar G(1,2)G(2,1),
\end{equation}
according to Eq.~\eqref{B:L0}. Then the correlation contribution
takes the form
\begin{equation}
K_{\mathrm{c}}(12,34) =  -
\delta(1-3)\delta(2-4) V(1,\bar{5}) 
L^{0}(\bar{5}\bar{5},\bar{6}\bar{6}) V(2,\bar{6})
\label{Kc10}
\end{equation}
The diagrammatic representation of Eq.~\eqref{L-K} with
kernels~\eqref{K-HF} and \eqref{Kc10} is shown in
Fig.~\ref{fig:BSE}. The correlation contribution in this
approximation is due to particle-hole pair excitation
processes~\cite{Baym-Kadanoff:PR,Kwong-Bonitz}.

To make the connection with the BS equation with screened-exchange
kernel, one can combine the $K_{\mathrm{HF}}$ and $K_{\mathrm{c}}$
contributions  to obtain
\begin{equation}
L(12,34)= L^0(12,34)
 +
L^0(12,\bar{5}\bar{6})K_{\mathrm{SX}}(\bar{5}\bar{6},\bar{7}\bar{8})
L(\bar{7}\bar{8},34) \label{L-KSX}
\end{equation}
where
\begin{align}
K_{\mathrm{SX}}(12,34) =\mbox{} &\delta(1-2)\delta(3-4) V(1,3)
- W(12,34) \label{KSX} \\
W(12,34) = \mbox{}& \delta(1-3)\delta(2-4) [V(1,2) + V(1,\bar{5}) L^{0}(\bar{5}\bar{5},\bar{6}\bar{6}) V(\bar{6},2)
].\label{B:W4}
\end{align}
This way, a general BS equation with screened-ex\-change kernel and
dielectric function
\begin{equation}
\varepsilon^{-1}(1,2)= \delta(1-2) +V(1,\bar{3}) \chi^{0}(\bar{3},2)
\end{equation}
is recovered.  However, there is an important difference between the
approach presented here and GW-based BS equation calculations:
Usually the BS equation is employ\-ed \emph{to determine the
absorption} using the screened potential as an ingredient. We
compute the density-density correlation function $\chi$, which is
related to $L$ via Eq.~\eqref{chi-L}, \emph{in order to determine
the screened potential $W$}.  Thus the meaningful comparison should
be made between RPA-like inverse dieletric functions, and the
$\varepsilon^{-1}$ resulting from the density-density correlation
function, or Eq.~\eqref{L-K}. As already mentioned in connection
with Eq.~\eqref{W-RPA}, retaining only the Hartree term in
Eq.~\eqref{L-K} is already equivalent to the
RPA~\cite{Baym-Kadanoff:PR,Kwong-Bonitz}, so that the inclusion of
exchange and scattering/dephasing contributions represents a
considerable improvement over the RPA. In addition, this improvement
is reached in a consistent manner, i.e., we can be reasonably sure
that we have not sacrificed accuracy at a different level in the
calculation.

Finally, we would like to relate Eq.~(\ref{L-K}) together with
\eqref{chi-L} to our derivation of the central Eq.~(\ref{eqnmotchi})
for $\chi^{\mathrm r}$ using nonequilibrium Green's functions
techniques. To facilitate the comparison, we perform a partial
resummation of Eq.~\eqref{L-K}. To display the HF contribution at
the two-particle level we define
\begin{equation}
L_{\mathrm{HF}}(12,34) = L^0(12,34) 
+ L^0(12,\bar{5}\bar{6}) 
K_{\mathrm{HF}}(\bar{5}\bar{6},\bar{7}\bar{8})
L_{\mathrm{HF}}(\bar{7}\bar{8},34) .
\end{equation}
This allows one to rewrite Eq.~\eqref{L-K} in the form
\begin{equation}
 L(12,34) = L_{\mathrm{HF}}(12,34) 
+ L_{\mathrm{HF}} (12,\bar{5}\bar{6}) K_{\mathrm{c}}
(\bar{5}\bar{6},\bar{7}\bar{8})
L(\bar{7}\bar{8},34) , \label{LHF-K}
\end{equation}
If one demands consistency also at the HF level then it is important
that the HF Eq.~\eqref{LHF-K} for the four-point functions use $L^0$
defined in Eq.~\eqref{B:L0} constructed from HF single-particle
Green's functions.

Equation~(\ref{LHF-K}) is still defined for time arguments on the
Keldysh contour, but it shows a formal similarity to
Eq.~\eqref{eqnmotchi} in that it consistently displays the HF
contributions on the one-particle and two-particle levels, and
separates out the correlation contributions, as it is also evident
in Eq.~\eqref{eqnmotchi}. To make use of Eq.~\eqref{LHF-K}, one has
to compute all the different Keldysh components of these four-point
quantities~\cite{4index}. Only after the full four-point correlation
function $L$ is calculated, one could obtain via Eq.~\eqref{chi-L}
the retarded density-density correlation function needed to
determine the screened Coulomb potential for the GW calculation. Our
derivation using kinetic equations and the generalized Kadanoff-Baym
ansatz, on the other hand, is a way to derive a closed set of
equations that depends only on two-time quantities \emph{and} obeys
the consistency condition derived from density/current conservation.
The generalized Kadanoff-Baym ansatz makes the calculational
procedure in principle self-consistent because it connects the
kinetic Green's functions in the correlation contributions with the
retarded (equilibrium) Green's functions. It also ensures that the
calculation deals exclusively with two-time quantities, which in
equilibrium depend only on the time difference, so that after
Fourier transformation both the one-particle and two-particle
correlation functions used in the calculation can be given a
quasiparticle interpretation.


\begin{thebibliography}{41}
\bibitem{4index}
\bibinfo{note}{Note that it is hard to give correlation functions depending on
  four Keldysh indices a direct physical meaning, as it is possible for, say,
  the retarded and kinetic components of Green's functions with two Keldysh
  indices; see, e.g., Ref.~\cite{Kremp:Book}.}
\end{thebibliography}
\end{document}